\documentclass{aa}  
\usepackage{graphicx}
\usepackage{txfonts}
\usepackage{hyperref}
\begin{document}

   \title{Influence of misalignments on performance of externally occulted solar coronagraphs}

   \subtitle{Application to PROBA-3/ASPIICS}

   \author{S.~V.~Shestov\inst{1}\fnmsep\inst{2}  \and  A.~N.~Zhukov\inst{1}\fnmsep\inst{3}
          }

   \institute{Solar-Terrestrial Centre of Excellence -- SIDC, Royal Observatory of Belgium, \\
              Avenue Circulaire 3, B-1180, Brussels, Belgium;  \email{s.shestov@oma.be}
	  \and
             Lebedev Physical Institute, Leninskii prospekt, 53, 119991, Moscow, Russia \\
	  \and
	     Skobeltsyn Institute of Nuclear Physics, Moscow State University, Leninskie gory, 119991, Moscow, Russia \\
             }

   \date{Received November XX, 2017; accepted Bombomber YY, 2017}
   \abstract
      {ASPIICS is a novel externally occulted coronagraph that will be launched onboard the PROBA-3 mission of the European Space Agency. The
	external occulter will be placed on one satellite $\sim 150$~m ahead of the second satellite that will carry an optical
	instrument. During 6 hours out of 19.38 hours of orbit, the satellites will fly in a precise (accuracy around a few millimetres) formation,
	constituting a giant externally occulted coronagraph. Large distance between the external occulter and the primary objective will
	allow observations of the white-light solar corona starting from extremely low heights $\sim 1.1R_\sun$.}
      {To analyze influence of shifts of the satellites and misalignments of optical elements on the ASPIICS performance in terms of
	diffracted light. Based on the quantitative influence of misalignments on diffracted light, we will provide a ``recipe'' for
	choosing the size of the internal occulter (IO) to achieve a trade-off between the minimal height of observations and sustainability
	to possible misalignments.} 
      {We consider different types of misalignments and analyze their influence from optical and computational points of view. We
	implement a numerical model of the diffracted light and its propagation through the optical system, and compute intensities of
	diffracted light throughout the instrument.  Our numerical model is based on the model of~\citet{2017A&A...599A...2R}, who
	considered the axi-symmetrical case. Here we extend the model to include non-symmetrical cases and possible misalignments.}
      {The numerical computations fully confirm main properties of the diffracted light that we obtained from semi-analytical consideration.
	We obtain that relative influences of various misalignments are significantly different. We show that the internal occulter with
	$R_{IO}=1.694$~mm $=1.1R_\sun$ is large enough to compensate possible misalignments expected to occur in PROBA-3/ASPIICS. Beside
	that we show that apodizing the edge of the internal occulter leads to additional suppression of the diffracted light.}
      {We conclude that the most important misalignment is the tilt of the telescope with respect to the line connecting the center of the
	external occulter and the entrance aperture. Special care should be taken to co-align the external occulter and the coronagraph,
	which means co-aligning the diffraction fringe from the external occulter and the internal occulter. We suggest that the best
      orientation strategy is to point the coronagraph to the center of the external occulter.}
   \maketitle

\section{Introduction}
  \label{intro}
  Solar corona is the outer layer of the solar atmosphere. It consists of highly ionized plasma that is structured by the magnetic field. In
  the solar corona a number of important but still ill-understood phenomena take place, such as the initiation and acceleration of coronal mass
  ejections (CMEs) and the acceleration of the solar wind.

  Nowadays the corona is observed by space-borne telescopes and coronagraphs in various spectral ranges: from X-rays and extreme ultraviolet
  (EUV) to white light. The EUV spectral range is more suitable for observations of the low corona due to extremely small brightness of plasma
  in EUV at high altitudes that decreases with height proportionally to the square of the electron density $n_e^2$, contrary to white-light
  brightness proportional to $n_e$. There are occasional observations up to $2 R_\sun$ with regular EUV telescopes that operated in
  dedicated long-exposure regimes,  e.g.  TESIS/CORONAS-PHOTON \citep{Kuzin2011,Reva2014} or SWAP/PROBA-2
  \citep{2006AdSpR..38.1807B,Seaton2013,2017JSWSC...7A...7D}, or up to $2.5 R_\sun$ with specialized EUV coronagraphs, e.g.
  SPIRIT/CORONAS-F \citep{2003AdSpR..32..473Z,2008AnGeo..26.3007S}.

  White-light coronagraphs can provide reliable observations over a significantly larger field of view due to a different mechanism of
  emission, primarily Thomson scattering. For example, the field of view of the internally occulted coronagraph COR1/STEREO ranges from
  $1.5$ to $4.0R_\sun$ \citep{Howard2008}. The field of view of externally occulted coronagraphs usually starts slightly higher, from $\sim
  2.2R_\sun$ in LASCO C2/SOHO \citep{Brueckner1995}, and from $2.5R_\sun$ in COR2/STEREO-A \citep{Howard2008} \citep[see also comparison of
  various coronagraphs in][]{Frazin2012}. The very important range of heights from $1.1R_\sun$ to $2.5R_\sun$
  \citep{2008ApJ...680.1532Z} is almost not covered in  white-light observations from space, except for LASCO C1/SOHO $\sim
  1.1-3.0R_\sun$ \citep{Brueckner1995}, which suffered from significant stray light.

  Classical internally occulted coronagraphs can not provide such observations as the theoretical limit of diffracted light around $10^{-4} B_\sun$ at
  $1.1R_\sun$ exceeds the brightness of the solar corona $10^{-5}B_\sun$ \citep[see review of various coronagraph systems in][hereafter
  RR17]{2017A&A...599A...2R}. Scattering of the bright light from the solar disk in the primary objective further worsens the problem. 
  The intensity of diffracted light in the inner corona region can be decreased by the use of apodized entrance aperture
  \citep{2007A&A...467..317A}, but this solution was not yet implemented in a solar coronagraph.
  
  For externally occulted coronagraphs, there are two fundamental factors that complicate observations at low heights: these are drastic
  increase of diffracted light and vignetting of the inner field of view \citep[see e.g.][]{Koutchmy1988,JGRA:JGRA53203}. Furthermore,
  straightforward ways to compensate both effects are contradictory: diffracted light decreases \citep{Lenskii1988,Fort1978}, whereas
  vignetting increases with increase of the external occulter size (or putting it closer to the primary objective). In all the previous
  space-borne coronagraphs, the distance from the external occulter to the primary objective was limited by the length of the instrument,
  i.e. $\sim 1$~m. This fact does not allow using sufficiently large primary objectives and simultaneously results in almost 50\%
  vignetting of the objective throughout the field of view.
  
  ASPIICS (Association of Spacecraft for Polarimetric and Imaging Investigation of the Corona of the Sun) is a novel white-light coronagraph
  \citep{2010SPIE.7731E..18L, 2015SPIE.9604E..0AR,Renotte2016}, that will perform regular observations of the corona over the field of view
  from $\sim 1.08R_\sun$ up to $3.0 R_\sun$. This will be possible thanks to European Space Agency PROBA-3 mission that will use formation
  flying (FF).  Such a technology allows virtual enlarging of instrumentation to an unprecedented size: the external occulter will be placed
  on one satellite and the optical instrument -- on the second. Both satellites  will synchronously fly on a highly elliptical 19.38-hour
  orbit and form a precise formation with inter-satellite distance $\sim 150$~m during 6 hours. The external occulter with $\diameter
  1.42$~m diameter will produce $\diameter77$~mm shadow, and the telescope with the entrance aperture $\diameter 50$~mm will be placed 
  in the center with the maximal possible precision. The giant inter-satellite distance allows unvignetted observations of the solar corona
  starting already from $\approx 1.1R_\sun$. 
  
  In order to provide highest possible accuracy of FF, the satellites are equipped with fine metrology subsystems that include laser
  beam/reflector and shadow position sensor (SPS) \citep{Renotte2016,2015SPIE.9604E..0CB}. SPS consists of 8 photodiodes, placed around the entrance aperture
  of the telescope in the penumbra region. SPS will measure relative position of the satellites with the accuracy better than $\sim 1$~mm in
  longitudinal and transversal directions. The expected FF accuracies are: $\pm 15$~mm in longitudinal and $\pm 5$~mm in transversal $x$-,
  and $y$-directions, $30$~arcmin for the attitude of the occulter satellite, and $15$~arcsec for the attitude of the satellite with the
  telescope ($3\sigma$ values). 
  
  Various analyses \citep{Landini2010,2013A&A...558A.138A} show that the amount of diffracted light reaching the ASPIICS primary objective
  is high enough. In order to reduce its level, the external occulter of ASPIICS will have a toroidal shape; technological limitations do not
  allow using a superior external occulter like a multi-disk system or a threaded-cone \citep[see][for review of various systems]{Bout2000}.
  RR17 used analytical/numerical approach and calculated not only the diffraction level just in front of the primary
  objective, but, more importantly, the spatial distribution of diffracted light at the detector. Authors found that the intensity drastically
  depends on the sizes of the internal occulter and the Lyot stop, and provided a ``recipe'' for choosing their proper sizes. The analysis,
  however, was limited only to the symmetrical case, when the Sun, the external occulter and the coronagraph are on the same axis and are
  perfectly co-aligned. 
  
  Beside the misalignments introduced by FF, there are additional sources of misalignments expected in ASPIICS. These can be  long- or
  short-term thermal expansions of mechanical structure, errors of initial co-alignment of the telescope on the satellite etc. In total these
  misalignments can be very small, of the order of 10--30~arcsec. Nevertheless, as \citet{Venet2010} concluded, even such low values can be
  critical for coronagraphs with extremely small overoccultation. 

  The aim of the present paper is to investigate how different misalignments influence the intensity of the diffracted light in the final image,
  to compare it with the intensity of the corona and to choose the proper size of the internal occulter to ensure reliable rejection for the cases of
  possible misalignments. The analysis considers only the effect of diffraction, i.e. the effect that is caused by masking of individual
  parts of the wavefront of propagating light. Other effects that will unavoidably be present and degrade the performance of the coronagraph --
  scattered light and ghosts, non-ideal lenses, etc., will provide an additional contribution but are not considered here. Whereas the
  computations are performed for a particular geometry and optical layout (representative for the ASPIICS coronagraph), we note that the
  obtained properties of the diffracted light and its behaviour with various misalignments remain valid for any externally occulted coronagraph. 

  The paper is structured as follows: in Sect.~\ref{sec2} we describe the optical layout and the basics of the algorithm, in Sect.~\ref{properties} we
  discuss some properties of the diffracted light, in Sect.~\ref{types-sec} we consider possible types of misalignments. In
  Sect.~\ref{results} we present results of computations, and consider different sizes of IO and its apodization. In Sect.~\ref{discussion-sec} we
  discuss the obtained results, and give conclusions in Sect.~\ref{conclusion-sec}. Appendix~\ref{method} contains mathematical details of
  the calculations, Appendix~\ref{numerical} details the numerical computations, and Appendix~\ref{sampling-sec} analyzes influence of
  samplings.

\section{Optical layout, model, algorithm}
\label{sec2}
\subsection{Optical layout}
  \label{layout-sec}
  A detailed description of the optical layout of ASPIICS is given by \citet{2015SPIE.9604E..0BG}. In the present paper we use a simplified model
  and conventions to denote optical planes described in RR17. Table~\ref{planes-table} summarizes the names and the description of the optical
  planes. 
  
  \begin{table}
    \caption{Key planes of the ASPIICS coronagraph}
    \label{planes-table}
    \begin{tabular}{l l}
      \hline \hline
      Plane & Description \\
      \hline
      $O$   & External occulter plane \\
      $A$   & Entrance aperture of the telescope \\
      $B$   & Focal plane of the primary objective  \\
      $O'$  & Conjugate image of plane $O$ (plane of the IO) \\
      $C$   & Conjugate image of plane $A$ (plane of the Lyot stop) \\
      $D$   & Final focal plane and the detector \\
      \hline
    \end{tabular}
  \end{table}
   
\begin{figure*}
    \centering
    \includegraphics[width=18cm]{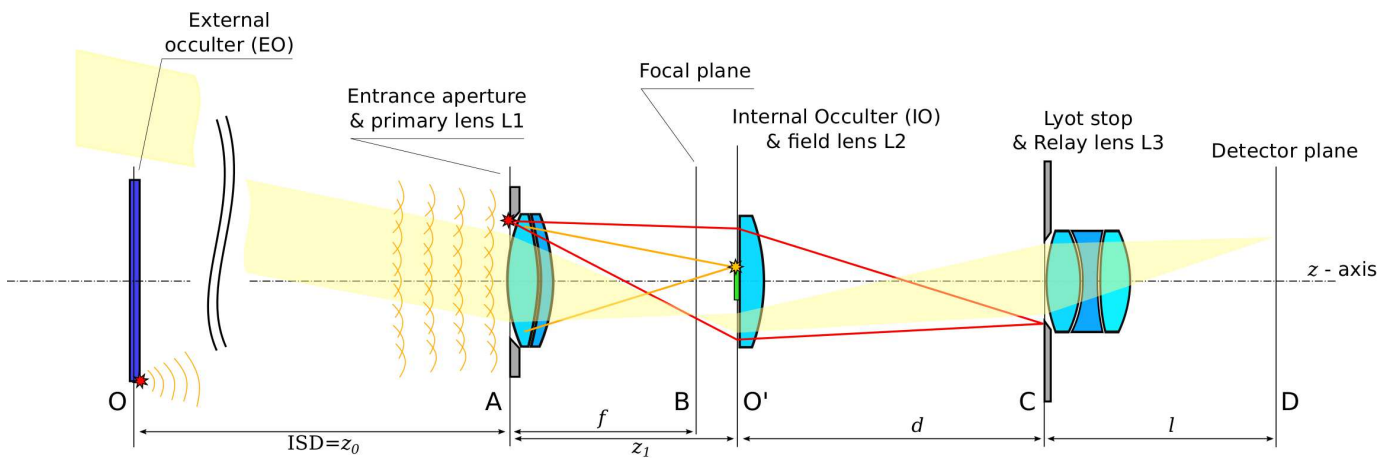}
    \caption{Optical layout of the ASPIICS coronagraph. Dim yellow light represents a regular coronal beam, orange
	    arches and lines represent the light that diffracts on the EO, the red lines represent the light diffracted at the entrance aperture.}
    \label{layout}
  \end{figure*}

  The optical layout of the ASPIICS coronagraph is given in Fig.~\ref{layout}.  The external occulter --
  EO (plane $O$) is situated on the occulter satellite placed at the distance $z_0=144\,348$~mm ahead of the coronagraph satellite with the
  optical instrument. The telescope consists of the entrance aperture (plane $A$) and the primary lens L1 with the focal length $f$, that makes
  an image of the corona in the primary focal plane $B$.  The light diffracted at the EO is focused in the $O'$ plane, situated at the distance
  $z_1$. This is the plane of the conjugate image of the EO produced by L1, thus it is $\Delta =z_1-f \approx 0.76$~mm further than $B$. The major
  part of the diffracted light is cut by the internal occulter, the size of which should be carefully chosen to obtain a good rejection of
  diffraction and not to vignette too much. The IO is deposited directly on the surface of the field lens L2. The lens L2 along with L1 make
  an image of the entrance aperture on the $C$ plane. The Lyot stop is placed in this plane and rejects the light diffracted at the
  entrance aperture. Simultaneously, the lens L2 projects the entrance aperture to the relay lens L3, ensuring that the light propagates
  further into the coronagraph and justifying its name -- the field lens.  Finally, both L2 and L3 project the primary focus $B$ onto the
  detector plane $D$. 

  In Fig.~\ref{layout} the dim yellow light represents a regular coronal beam, orange arches represent light that diffracts on the EO. After
  propagating through the entrance aperture and L1, this light is focused (orange lines) in the $O'$ plane. The red lines represent the
  light diffracted at the entrance aperture.

  Our optical model contains two main simplifications: we place the entrance aperture and the primary objective into the same plane $A$ (in the real
  instrument the objective is $\sim 120$~mm behind), and we change the distances $d$, $l$ and the focal distance of L3 to keep the same spatial scale
  in $D$ as in $B$. Both simplifications do not influence the diffracted light.  

  We perform the calculation for the central wavelength 550~nm of the main (wide-band) passband and discuss possible effects in the
   Sect.~\ref{wavelength}. 

  Table~\ref{geometry-table} summarizes main parameters of the ASPIICS optical layout and other parameters used in our computations.

  We consider an infinitely thin (razor edge) external occulter. In reality the edge has a toroidal cross-section both for improvement
  of the diffracted light rejection \citep{Landini2010}, and for reduction of the effect of tilting. The simplification used here is linked
  to the difficulty of a correct representation of 3D occulters in the framework of the Fresnel diffraction \citep{Sirbu:16}. Currently
  attempts are ongoing to take the effect of the toroidal shape of the occulter into account both, experimentally and numerically.

  \begin{table}
    \caption{Parameters used in our numerical study taken from the ASPIICS configuration. See text for details.}
    \label{geometry-table}
    \begin{tabular}{l l l}
      \hline \hline 
      Parameter & Symb. & Value \\
      \hline
      Wavelength                      & $\lambda$ & $550$~nm \\
      Angular radius of the Sun	      & $R_\sun$ & 16~arcmin \\
      Radius of the EO                & $R_{EO}$  & 710~mm \\
      Distance plane $O$ -- plane $A$ & $z_0$     & $144\,348$~mm \\
      Angular radius of the EO	      & $\omega_{EO}$ & 16.909~arcmin \\
      Radius of the entrance aperture & $R_{A}$   & 25.0~mm \\
      Focal length of L1              & $f$       & 330.348~mm \\
      Distance plane $A$ -- plane $O'$ & $z_1$    & 331.143~mm \\
      Radius of the internal occulter  & $R_{IO}$  & 1.662~mm\tablefootmark{a} \\
      Hole in the internal occulter\tablefootmark{b} & $r_{IO}$  & 0.489~mm \\
      Lyot stop radius		      & $R_C$     & $0.97R_A$ (24.25~mm)\tablefootmark{c} \\
      \hline
    \end{tabular}
    \tablefoot{
      \tablefoottext{a}{the value that we use as a baseline for the diffracted light;}
      \tablefoottext{b}{a hole in the internal occulter is used to produce images of LEDs from the occulter satellite on the detector;}
      \tablefoottext{c}{it is the relative size of the entrance aperture and the Lyot stop that plays a role.}
    }
  \end{table}

\subsection{Projection of the Sun, EO and IO onto different planes}
  It is interesting to compare how different objects -- the Sun, the EO and the IO are projected in the framework of the geometrical optics
  onto planes $B$ and $O'$. It turns out that the geometrical approach correctly predicts the relative sizes of images, their relative shifts due to
  the tilt of the coronagraph or other misalignments. Since $O$ and $O'$ are conjugate planes, we can project the IO onto $O$ and after that onto $B$.
  In the figures below we show the EO and IO with a transparent central part for clarity. 
  
\begin{figure}
    \resizebox{\hsize}{!}{\includegraphics{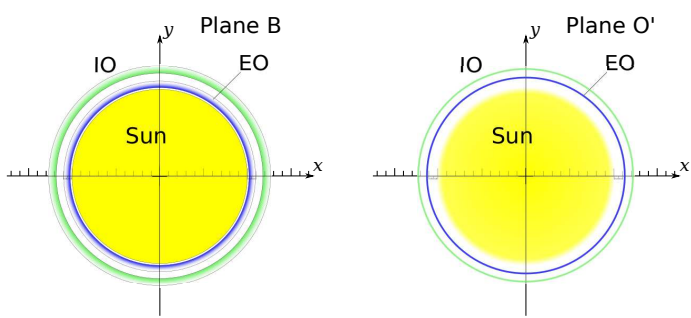}}
    \caption{Projections of the Sun, the EO and IO into planes $B$ and $O'$ in symmetrical case.}
    \label{symmetrical}
  \end{figure}

  We start with the case of the perfect symmetry, when the coronagraph and the EO are co-aligned and co-centered, and the center of the Sun
  lies on the same optical axis $z$. In Fig.~\ref{symmetrical} we present sketches with plane $B$ in the left panel and plane $O'$ in the
  right panel.
  
  The Sun is sharply focused in the $B$ plane, and slightly defocused in the $O'$ plane (we denote defocusing by a smooth limb). 
  
  The EO is represented by the blue color in both planes. In the $O'$ plane the image of EO is perfectly focused, which is denoted by a thin
  ring. In the $B$ plane the image of the EO is defocused, thus the EO is marked as a ring with a finite thickness and gradient filling.
  Inner and outer edges of the ring correspond to the extremes in optical vignetting produced by the EO, and the gradient itself represents
  the vignetting function (full obscuration inside and full transparency outside).  Obviously, the angular size of the EO should be selected
  in such a way that its inner vignetting zone is larger than the Sun. 

  The IO is represented by a green ring. Similarly to the EO, it is perfectly focused in the $O'$ plane and defocused in the $B$ plane. The
  reasoning about the inner and the outer edges can be applied here as well. The size of the IO should be selected in such a way that it fully covers
  both the Sun and the EO. 

\subsection{Algorithm of the diffraction calculation}
  \label{algorithm}
    In our analysis we follow the basic idea of \citet{2013A&A...558A.138A} and RR17. Here we briefly summarize the method, and we give
    the full mathematical description in Appendix~\ref{method}.
    
    We setup a reference frame with the $xy$ plane coinciding with the $O$ plane and co-centered with the EO, and the $z$-axis pointing to the
    entrance aperture (this implies the co-alignment and co-centering of the coronagraph and the EO). The Sun is considered as a spatially
    extended light source that produces a set of non-coherent plane-parallel waves with wave-vectors $\vec k$:
    \begin{equation}
      \Psi_{\vec k} = A \exp { \left(-i (\omega t - \vec k \vec r) \right) }.
      \label{plane-parallel}
    \end{equation}
    By varying the direction of $\vec k$ we sample the solar disk (so far it needs not to be co-aligned).

    The algorithm for calculating the diffracted light in various planes consists of the following:
    \begin{enumerate}
      \item We select a particular point on the Sun and consider its plane-parallel wave $\Psi_{\vec k}$. The direction of the wave-vector
	$\vec k$ can be specified in the $B$ plane either by Cartesian $(\alpha,\beta)$, or by polar $(\rho,\varphi)$ coordinates. The
	intensity of the wave depends on the solar brightness and the sampling area: $A^2 = B_{\vec k} \mbox{d}S$.
      \item After diffraction at the EO, each wave becomes significantly not a plane-parallel one, and it can not be expressed similarly to
	(\ref{plane-parallel}). We express its amplitude in the $A$ plane as $\Psi_{A \vec k}$, where $\vec k$ denotes the orientation of
	initial wave. 
      \item Further propagation of the wave is considered in the framework of the Fourier optics formalism. The fields in the $O'$,
	$C$, and $D$ planes are calculated with three successive Fourier-Fresnel transforms over the distances $z_1$, $d$, $l$ and taking
	into account apertures $A_A$, $A_{O'}$, and $A_C$.
      \item We consider all the possible waves, summarizing over directions of $\vec k$ (i.e. over $(\rho,\varphi)$ or
        $(\alpha,\beta)$), and calculate the final image in $D$ as:
	\begin{equation}
	  I_D = \sum_{\vec k} B_{\vec k} |\Psi_{D \vec k}|^2 \, \mbox{d}S.
        \end{equation}
    \end{enumerate}

    We stress that the $\vec k$ vector is determined in the reference frame of the EO and the coronagraph, not that of the
    Sun. All the characteristics of the Sun -- the solar brightness, angular size, limb darkening -- are implicitly hidden in $B_{\vec k}$
    and possible $\vec k$ directions, whereas the obscuration by the EO and IO is implicitly hidden in $\Psi$. 
    
\begin{figure}
      \resizebox{\hsize}{!}{\includegraphics{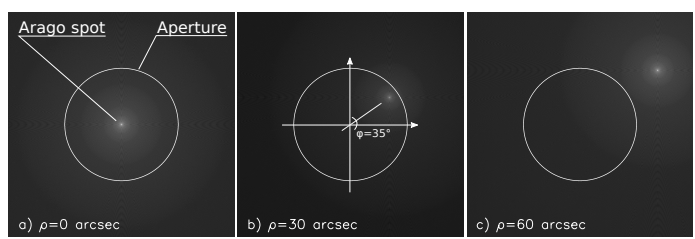}}
      \caption{Intensities $|\Psi_{A \vec k}|^2$ of waves in the $A$ plane propagating with different directions of $\vec k_{\rho\varphi}$: a)
      $\rho=0$, b) $\rho=30\arcsec$, and c) $\rho=60\arcsec$ ($\varphi=35^\circ$ in the latter two cases). Only the part of $\Psi_A$ that enters into
      the aperture produces signal in further planes of the telescope.}
      \label{tilted-waves}
    \end{figure}

    Calculation of $\Psi_{A \vec k}$ is a separate problem as it involves computing of the Fresnel diffraction on the EO.  Again, we follow
    approach of \citet{2013A&A...558A.138A} and RR17, who considered an infinitely thin (razor edge) external occulter. The details are
    given in Appendix~\ref{method}, and here we note that $\Psi_{A \vec k}$ for a particular $\vec k$ can be calculated from the co-axial
    wave $\Psi_{A00}$ ($\rho=0, \, \varphi=0$) by shifting and multiplicating by an arbitrary phase function. We give several examples of
    $\Psi_{A \vec k}$ in Fig.~\ref{tilted-waves}, where the three panels correspond to different $\vec k$: the co-axial one with $\rho=0$
    (this is essentially $\Psi_{A00}$), and two tilted ones with $\rho=30\arcsec$ and $\rho=60\arcsec$ ($\varphi=35^\circ$ in the latter two
    cases). The central bright feature of the co-axial wave -- the Arago spot is displaced in tilted waves. Highly oscillating nature of the
    $\Psi_{A \vec k}$ function (see Fig.~2 in RR17) is present but is not visible here in the pdf/hardcopy images. Only the part of the wave
    that fits the entrance aperture produces further signal in the instrument. The full-Sun umbra/penumbra pattern in the aperture plane can
    be obtained by integrating $\Psi_{A\vec k}$ over the Sun and resembles geometrical umbra/penumbra pattern with some signal in the umbra
    region (see Fig.~5 in RR17).
    
\section{Some properties of diffraction}
  \label{properties}
  Calculation of the diffracted image in the $O'$ plane $I_{O'}=\sum_{\vec k} B_{\vec k} |\Psi_{O' \vec k}|^2 \mbox{d}S$ reveals that it is a
  bright ring with radius $r=z_1 \tan ( \omega_{EO})$, i.e. with the angular size \emph{exactly} coinciding with the angular size of the EO -- $\omega_{EO} =
  \arctan \left( R_{EO}/z_0 \right)$.  
  
\begin{figure*}
    \includegraphics{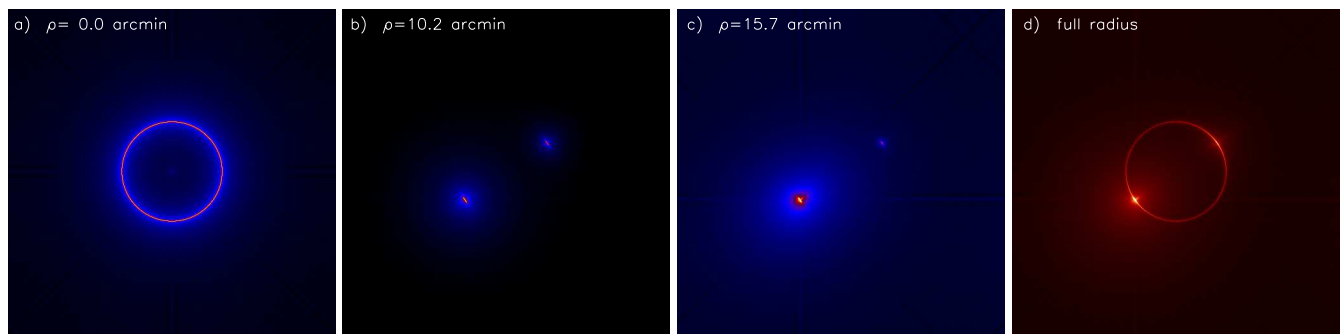}
    \caption{Intensity of the diffracted wave $|\Psi_{O'\vec k}|^2$ in the $O'$ plane for different initial waves $\vec k_{\rho\varphi}$: a)
    co-axial wave $\rho=0$, b) moderately tilted wave with $\rho=10.24\arcmin$, c) highly tilted wave with $\rho=16\arcmin$, d) integrated along the
    solar radius; $\mbox{d}S = \rho \, \mbox{d} \rho \, \mbox{d}\varphi$ was taken into account during integration, thus the color table is
    different; $\varphi=35^\circ$ was used in the latter three cases.}
    \label{property}
  \end{figure*}

  One of the main properties of the diffraction is the following: every single wave $\Psi_{A\vec k}$ after propagation
  into $O'$ highlights there either \emph{a sector} or \emph{the whole ring} (see Fig.~\ref{property} with examples of  $|\Psi_{O'\vec
  k}|^2$). The highlighted sector has the same polar coordinate $\varphi$ as the initial wave, and its angular width depends on
  $\rho$ of the initial wave: the higher $\rho$, the smaller the width.  For the co-axial wave $\rho=0$, and the whole ring is
  bright (panel a). For the waves with $\rho \approx 10 \arcmin$ and $\rho \approx 16\arcmin$ the bright sector becomes very
  narrow (panels b and c). Two-dimensional distributions of intensities have very intense peaks and strongly decreasing wings. 
  
  This reasoning allows inferring the second property: if the EO and the coronagraph are on the same optical axis, the relative
  displacement of the Sun from this axis \emph{does not} change the geometrical symmetry of the diffraction ring in $O'$. In the case of such
  a displacement, individual parts of the diffracted ring become brighter or dimmer, but they
  do not move in the $O'$ plane. Similarly, seasonal change of the angular size of the Sun \emph{does not} change the geometry of the
  diffracted ring in $O'$.

  The final image in $D$ is formed by diffracted light that ``propagates behind'' the IO. As long as the relative shift of the diffracted ring
  and the IO is small (as expected in ASPIICS), the effect of diffraction determines the final image in $D$. Thus, the third property is:
  the position of the diffraction ring on the detector, its size and shape are determined by the IO. Misalignments do not shift or distort
  the final diffraction ring; they just make individual parts of the ring brighter or dimmer. It is the variations of the IO size, that can change the size of the final
  diffraction ring. 

\section{Types of misalignments}
    \label{types-sec}
    There are various types of misalignments of optical elements that influence the overall intensity and spatial distribution of the
    diffracted light. Among them are transversal (in the $xy$-plane) shifts of the EO and the telescope, tilts of the EO and the telescope, shifts
    and tilts of internal optical elements inside the coronagraph, change of $z_0$, etc.  Influence of some of the misalignments, such as
    change of $z_0$ or the axial displacement of the IO, can be analysed using the symmetrical model since they do not break the axial symmetry.
    Other misalignments, such as the transversal shift of the EO and the coronagraph, or transversal shifts of apertures must be
    considered using full-sampling approach, as the optical layout becomes not axi-symmetrical.
    
    Below we will show that any configuration with transversal shifts of ``external'' optical elements can be reduced to superposition of
    just two of them -- tilt of the telescope and shift of the Sun. 
    
    We are not going to consider influence of the possible tilt of the lenses (the expected tilts of the order of 0.5--5~arcmin will be
    negligible) and the tilt of the EO (due to the reasons explained in Sect.~\ref{layout-sec}).

    \begin{figure*}
      \includegraphics{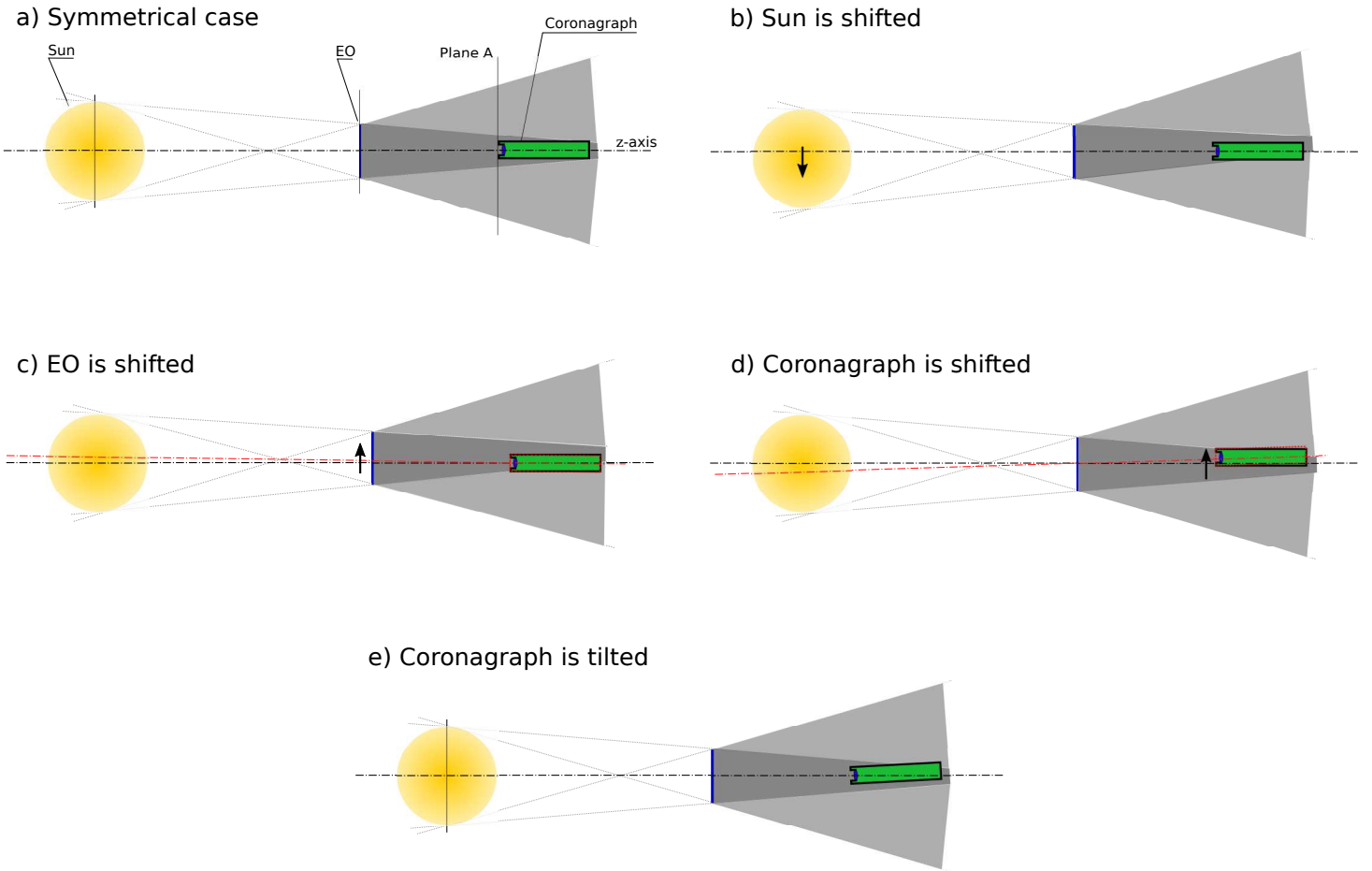}
      \caption{Comparison of various misalignments: a) symmetrical case; b) shift of the Sun; c) shift of the EO; d) shift of the coronagraph,
	and e) tilt of the coronagraph. In panels c) and d) the red axis goes through centers of the EO and the entrance aperture, and dotted
        rectangles represent the coronagraph co-aligned with this new axis.}
      \label{misalignments}
    \end{figure*}

    In the symmetrical case (panel (a) in Fig.~\ref{misalignments}) the $z$-axis goes through the center of the Sun, the EO and the coronagraph.
    It is perpendicular to the EO plane and co-aligned with the coronagraph optical axis.  The entrance aperture  is co-centered with
    the umbra/penumbra pattern.

\begin{figure}
      \resizebox{\hsize}{!}{\includegraphics{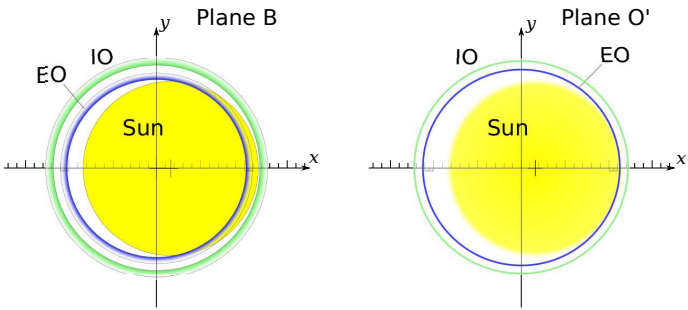}}
      \caption{Projections of the Sun, EO and IO into planes $B$ and $O'$ in case of a shifted Sun (cf. Fig.~\ref{symmetrical}).}
      \label{shift}
    \end{figure}

    In the case the Sun is shifted (panel (b) in Fig.~\ref{misalignments}), the EO and the coronagraph remain on the initial $z$-axis. However, the
    umbra/penumbra pattern is not symmetrical anymore with respect to the entrance aperture. The projections of the Sun, EO and IO into
    planes $B$ and $O'$ are shown in Fig.~\ref{shift}: the projection of the Sun is shifted, but the EO remains in the same position: the solar
    shift changes the relative intensity of EO (i.e. the diffraction bright ring), but does not change its geometry or its position (this is due to the property discussed in
    Sect.~\ref{properties}).  The coronagraph and its mathematical model remain axi-symmetrical, thus the waves propagating from different
    directions $\vec k_{\rho\varphi_1}$ and $\vec k_{\rho\varphi_2}$ produce \emph{similar} responses in $D$, which are just rotated with
    respect to each other by the angle $\varphi_2-\varphi_1$.  From the computational point of view the misalignment changes coordinates,
    e.g. $(\alpha,\beta) \rightarrow (\alpha+\phi,\beta)$, of all the waves $\Psi_{\vec k}$ that represent the Sun.

    The cases where the EO is shifted (panel c) or the coronagraph is shifted (panel d) are similar to each other. The umbra/penumbra pattern in
    the aperture plane becomes not symmetrical. We define a new axis (red dash-dotted line in Fig.~\ref{misalignments}) through the centers
    of the EO and the coronagraph. The new axis is tilted by the angle $\phi = \sqrt{\Delta x^2+\Delta y^2}/z_0 \le 10$~arcsec with respect to
    the original axis (black dash-dotted line) as the maximal expected shift is $\sqrt{\Delta x^2+\Delta y^2} \sim 7.0$~mm.  In the new reference
    frame we can neglect the tilt of the EO and take into account the tilt of the coronagraph and shift of the Sun (the co-aligned coronagraph is
    denoted by the red dashed line). Thus, these two cases can be considered as a simultaneous shift of the Sun and tilt of the coronagraph,
    with the two effects enhancing each other.
    
\begin{figure*}
      \centering
      \includegraphics[width=18cm]{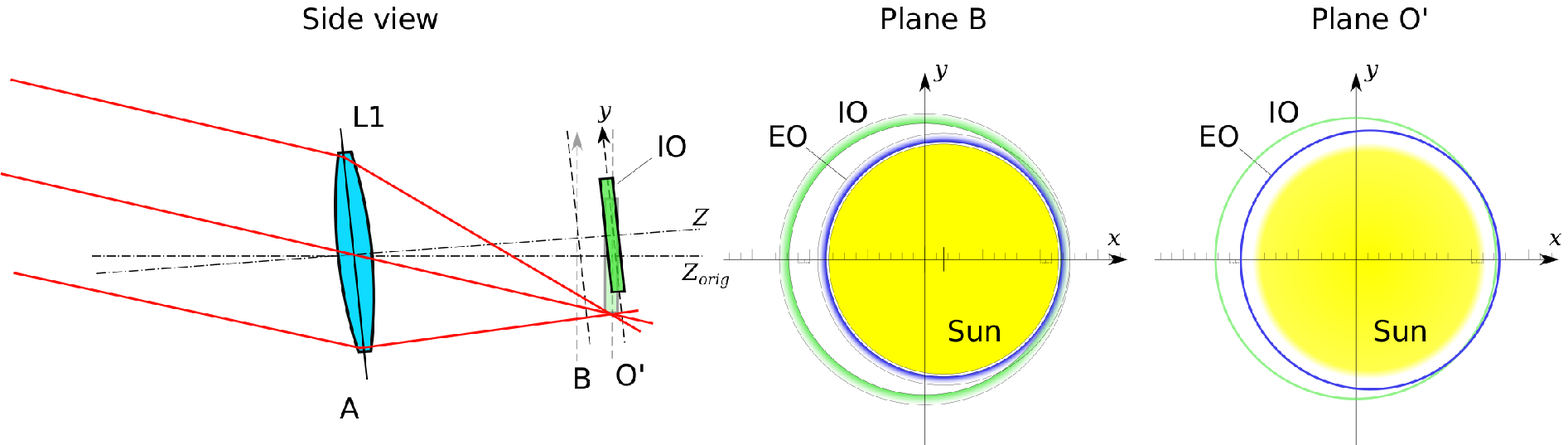}
      \caption{Tilt of the coronagraph. Left panel: relative displacement of the IO and the axis of the instruments. Middle and right panels:
	projections of the Sun, EO and IO onto planes $B$ and $O'$ (cf. Fig.~\ref{symmetrical} and Fig.~\ref{shift}).}
      \label{tilt}
    \end{figure*}

    In the case of the tilt of the coronagraph (panel e), the main effect is that in the $O'$ plane the projections of the Sun and the EO are displaced
    with respect to the IO (see Fig.~\ref{tilt}). Then more diffracted light propagates behind the IO on one side, and less -- on the other side.
    Another effect is the inclination of the EO image (and thus its defocussing), but is less important due to small expected values of the tilt $\sim
    25$~arcsec. Obviously, the overoccultion by the IO should be large enough to cover possible shifts of the EO image.  The coronagraph and
    its mathematical model become not axi-symmetrical: the two waves propagating from different directions $\vec k_{\rho\varphi_1}$ and
    $\vec k_{\rho\varphi_2}$ produce \emph{different} responses in $D$.  From the computational point of view, the tilt of the coronagraph
    around the $Oy$-axis is equivalent to multiplication of every $\Psi_{A \vec k}$ by the complex factor $\exp (-2 \pi i \frac{\Theta x}{\lambda} )$. 
    
    Transversal displacement of the IO is very similar to the case of the tilt of the coronagraph, so we do not present additional
    computations for it.  In the case of transversal displacement of the Lyot stop, the entrance aperture remains in the center of the umbra/penumbra pattern (it
    receives minimal possible level of the diffracted light) and the bright diffraction ring is co-centered with the IO (i.e. the IO acts with maximal
    efficiency). Thus we conclude that the influence of this misalignment will be smaller than that produced by the shift of the Sun and do not consider it
    separately.

    The case with the change of $z_0$ results in two effects: the change of level of the diffracted light on the entrance aperture,
    and change of the size of the diffraction ring in the $O'$ plane. The combined impact result of the two effects in the detector plane is not
    obvious and will be analyzed below. From the computational point of view, the misalignment requires recalculation of the
    axi-symmetrical function $\Psi_{A00}$ mentioned in Sect.~\ref{algorithm} (see also Appendix~\ref{numerical}).
    
    The effect of the longitudinal displacement of the IO from the $O'$ plane results in more diffracted light potentially propagate
    beyond the IO. From the computational point of view, it changes the $\Psi_{O'}$ function (see Appendix~\ref{method}).

\section{Results of computations}
  \label{results}  
  We consider $3\sigma$-values of the misalignments, resulting from the FF accuracy (see Sect.~\ref{intro}): relative shift of the Sun by
  $\phi=10$~arcsec (due to the transversal displacement of the satellites), tilt of the coronagraph by $\Theta \sim 10$ (due to the
  transversal displacement of the satellites), and 25~arcsec (due to the tilt of the satellite plus the misalignment of the telescope), change of
  $z_0$ by $+15$~mm and $-100$~mm (we exaggerate the expected $\Delta z_0 = -15$~mm to the extreme), longitudinal displacement of the IO by
  $\Delta z_{IO}=\pm 60$~$\mu$m.\footnote{Since the ASPIICS project is in the middle of the preparation phase, these values are subject to
  changes.}

  During the computations we use parameters  listed in Table~\ref{geometry-table}. We use $4k \times 4k$ arrays to represent $\Psi$ in
  each plane and sample the solar disk by $N=50$ points in radial direction and by $M=400$ points in polar direction~(see
  Appendix~\ref{sampling-sec} for the analysis of the sampling).

  We use the same solar limb darkening function as that used by RR17, which was initially proposed by \cite{1993AJ....106.2096V}:
  \begin{equation}
    B(\rho)=1-0.762 \left( 1-\sqrt{1-\rho^2} \right) -0.232(1-\rho^2)\log\left( \sqrt{1-\rho^2} \right),
  \end{equation}
  where $\rho$ is the radial coordinate expressed as a fraction of the solar radius.

  Since we are interested in the possibility of registering the signal by the ASPIICS detector, we convert the obtained brightness into
  number of photons pixel$^{-1}$. For the conversion we use the following ASPIICS parameters \citep{Renotte2016}: aperture size $A=\pi
  R_A^2=19.6$~cm$^2$, angular size of a pixel of $2.8$~arcsec, the exposure time $t_{exp}=0.1$~s, and take the mean solar brightness as
  $\mathrm{MSB}=2.08 \cdot 10^{20}$~photons s$^{-1}$ cm$^{-2}$ sr$^{-1}$ (the solar spectrum convolved with the spectral transmission
  of ASPIICS).

  To compare the diffracted light with the corona, we take the K-corona brightness observed during solar maximum from
  \citet{1977asqu.book.....A}. We take into account the geometrical vignetting, which is determined by the size of the IO: we apply linear
  vignetting function that starts (transmission $T=0\%$) from the angle $v_{min}=\arctan \left( \frac{R_{IO}^*-R_A}{z_0} \right)$, and
  finishes ($T=100\%$) at the angle $v_{max}=\arctan \left( \frac{R_{IO}^*+R_A}{z_0} \right)$ , where $R_{IO}^*$ stands for the size of the IO
  projected into $O$. In reality there will be an additional effect at low angles: high level of vignetting significantly
  widens the point-spread function of the coronagraph, which not only reduces the intensity, but also smears the image of the lower corona (see RR17
  for the analysis of the effect). 

  Further on we show profiles of the signal in the $D$ plane taken in the direction where the increase of the diffracted light is
  maximal, i.e. in the $Ox$ direction in $D$.

  \subsection{Qualitative results}
  Our numerical computations fully confirm the qualitative behaviour of images in different planes as described in Sect.~\ref{properties}. In
  the $O'$, plane the diffracted ring does not move due to solar shift, but it moves due to the tilt of the coronagraph. In the $D$ plane, the
  diffracted ring moves neither in the case of the solar shift, nor in the case of the tilt of the coronagraph.

  \subsection{Shift of the Sun and tilt of the coronagraph}  
\begin{figure}
      \resizebox{\hsize}{!}{\includegraphics{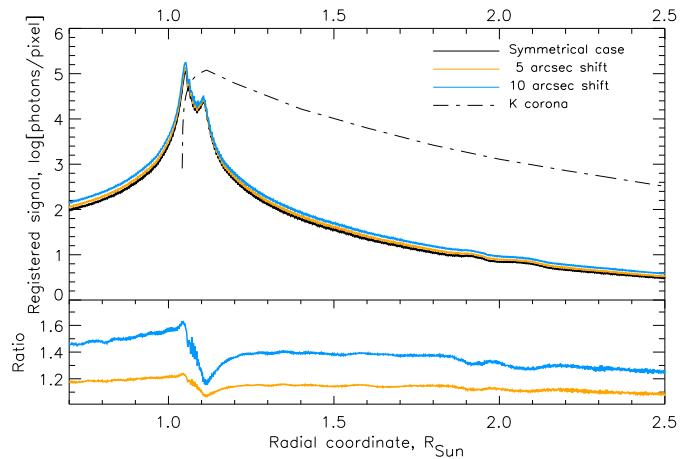}}
      \caption{Intensities of the diffracted light and the corona for the case of the shift of the Sun. Top panel: radial profiles computed
	for the symmetrical and shifted cases. The black curve denotes the symmetrical case, the yellow curve is for the 5~arcsec shift, the
	light blue curve is for the 10~arcsec shift. The dash-dotted line represents the brightness of the K-corona. Bottom panel: ratios of
	shifted and symmetrical profiles of the diffracted light.}
      \label{shift-profile}
    \end{figure}

    Intensities of the diffracted light for the case of the shift of the Sun are presented in Fig.~\ref{shift-profile}. In the top panel the radial
    profiles correspond to different values of shift: the black curve -- symmetrical (unshifted) case, the yellow curve -- $\phi=5$~arcsec,
    and the blue curve -- $\phi=10$~arcsec. The dash-dotted line shows the brightness of the K-corona. In the bottom panel, ratios of
    shifted and symmetrical profiles are given. 
  
\begin{figure}
      \resizebox{\hsize}{!}{\includegraphics{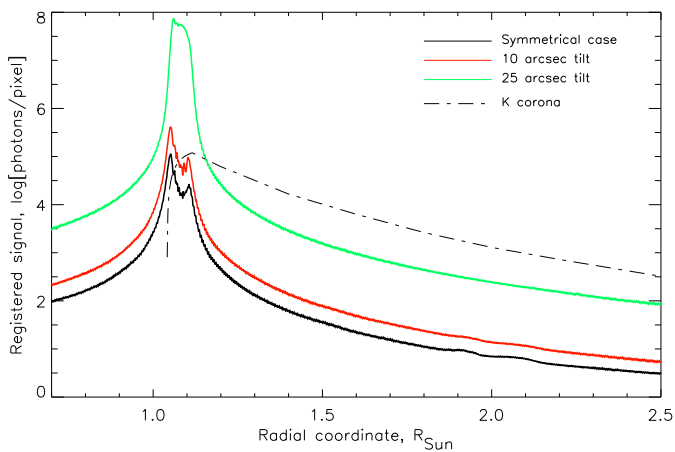}}
      \caption{Intensities of the diffracted light and the corona for the case of the tilt of the coronagraph. The black
	curve denotes symmetrical case, the red curve is for the 10~arcsec tilt, the green curve is for the 25~arcsec tilt. The dash-dotted line represents the
	brightness of the K-corona.}
      \label{tilt-profile}
    \end{figure}

    Intensities of the diffracted light for the case of the tilt of the coronagraph  are presented in Fig.~\ref{tilt-profile}. The black curve
    denotes the symmetrical case, the red curves is for the shift $\Theta=10$~arcsec, the green curve is for the $\Theta=25$~arcsec. The dash-dotted
    line shows the brightness of the K-corona.  As mentioned above, the radial position of the diffracted profile maximum does not shift
    with the tilt of the coronagraph. 
    
\begin{figure*}
      \includegraphics{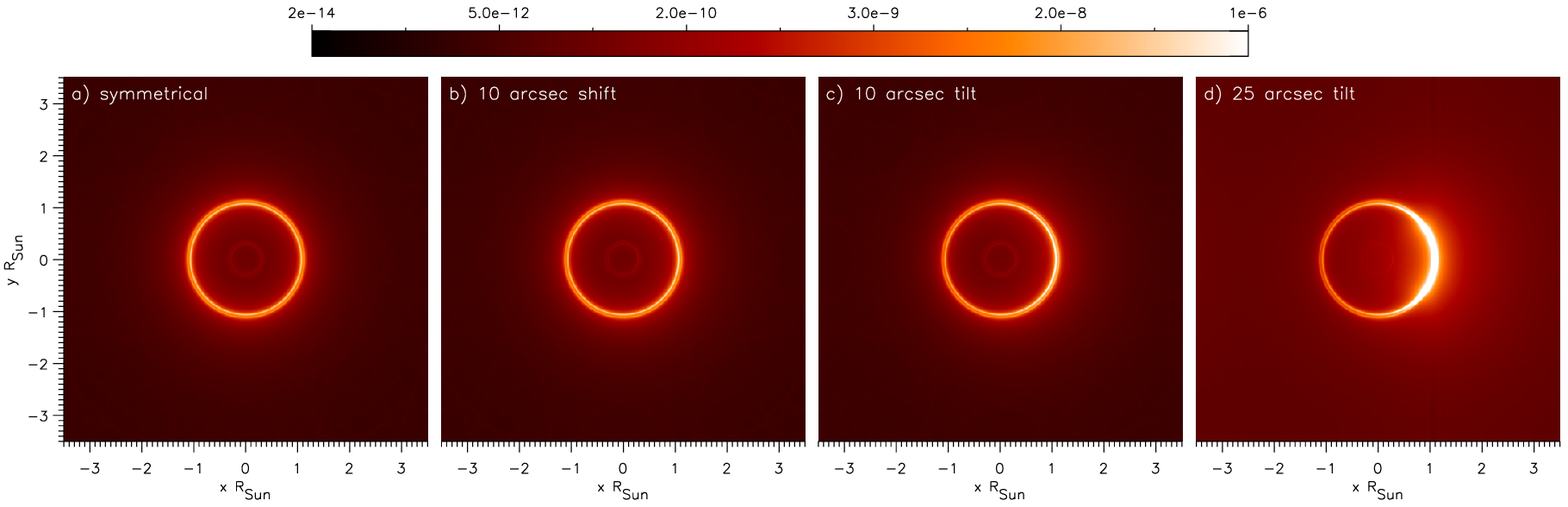}
      \caption{Images in the plane $D$ computed for: a) symmetrical case, b) solar shift by 10~arcsec, c) coronagraph tilt by
        10~arcsec, and d) coronagraph tilt by 25~arcsec. Color table is expressed in mean solar brightness.}
      \label{images}
    \end{figure*}

    In Fig.~\ref{images} we present images in the plane $D$ computed for symmetrical, shifted and tilted cases. All four panels are
    displayed with the same quasi-logarithmic color table. 

  \subsection{Longitudinal displacement of the IO and change of the inter-satellite distance $z_0$}
\begin{figure}
      \resizebox{\hsize}{!}{\includegraphics{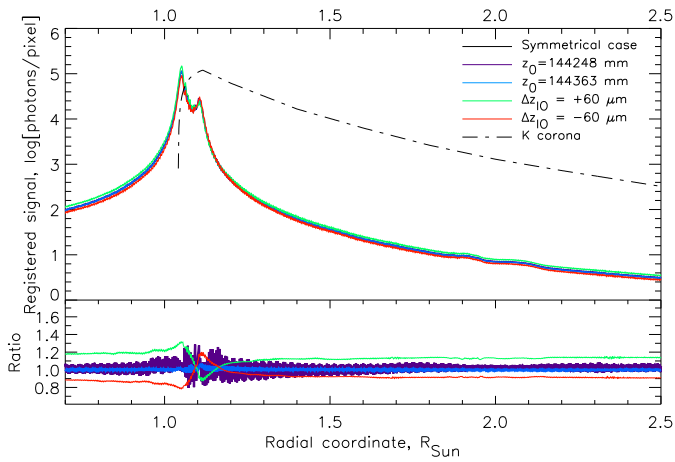}}
      \caption{Intensities of the diffracted light and the corona for the case of the longitudinal displacement of the IO
	and change of $z_0$. Top panel: radial profiles of the diffracted light (colored curves) and the K-corona (dash-dotted line). Bottom
	panel: ratios of the displaced and symmetrical cases.}
      \label{z1-z0-profile-ratio}
    \end{figure}

    The results for the relative displacements of the IO along the $z$-axis by $+60$~$\mu$m and $-60$~$\mu$m, and for the change of the inter-satellite
    distance $z_0$ by $+15$~mm and $-100$~mm are presented in Fig.~\ref{z1-z0-profile-ratio}. In the top panel, the radial profiles are given,
    in the bottom panel -- ratios of misaligned and symmetrical cases. All the changes are relatively small (up to 30\%), and can barely be seen
    in the top panel. Whereas the result of reducing $z_0$ by 100~mm is pronounced in the bottom plot, the changes are invisible in the top
    panel. This is probably due to oscillatory behavior of the obtained profile. The highest impact is produced by the longitudinal
    displacement of the IO by $+60$~$\mu$m further from L1, and the changes are comparable to those in the case of the shift of the Sun by 5~arcsec.
    The change of $z_0$ by $15$~mm and inward displacement of the IO almost do not change the overall intensity of the diffracted light.
    
  \subsection{Choosing the proper IO size}
    It is clear that the major impact on the diffracted light is produced by the tilt of the coronagraph and by the misalignments that can
    be reduced to such a tilt. Already the tilt of 25~arcsec is strong enough so that the diffracted light exceeds the level of the coronal
    intensity near the minimal heights. This is not unexpected if we consider projections of the EO and IO onto the $O'$ plane: the rightmost
    edge of the EO has the coordinates $x_{max}=z_1 \tan (25\arcsec + \omega_{EO})=1.669$~mm, which is larger than $R_{IO}=1.662$~mm. In
    other words, the bright diffraction ring is not fully blocked by the IO (this situation is shown in Fig.~\ref{tilt}). Whereas
    the difference of $7$~$\mu$m is small, a lot of diffracted light propagates further in the instrument.

\begin{figure}
      \resizebox{\hsize}{!}{\includegraphics{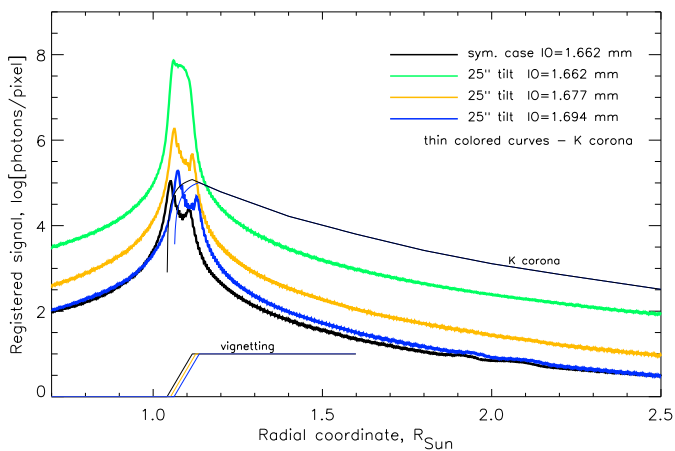}}
      \caption{Comparison of the diffracted light corresponding to different IO sizes for the same tilt of 25~arcsec. The green curve
	corresponds to $R_{IO}=1.662$~mm, the yellow curve is for the $R_{IO}=1.677$~mm, the blue curve is for the $R_{IO}=1.694$~mm, the
	black solid curve denotes symmetrical case with $R_{IO}=1.662$~mm. Thin black and blue lines denote the coronal signal with
      vignetting taken into account or the vignetting functions (in the bottom part of the plot.}
      \label{tilted-IOs}
    \end{figure}

    The impact of the tilt can be reduced if we increase the size of the IO. In Fig.~\ref{tilted-IOs} we compare the diffracted light for
    various sizes: $R_{IO}=1.662$~mm, 1.677~mm, and 1.694~mm for the same tilt of 25~arcsec.  Already $R_{IO}=1.677$~mm reduces the diffracted
    light by an order of magnitude. However, the diffracted light is still high enough at heights $<1.15 R_\sun$. The IO with $R_{IO}=1.694$~mm
    almost completely removes the effect of the tilt, as the intensity of the diffracted light becomes almost equal to the symmetrical case with
    $R_{IO}=1.662$~mm. We note, however, that due to the increase of the IO both the peak of the diffracted light intensity and the coronal vignetting
    shift towards higher altitudes with respect to the symmetrical configuration. In the bottom part of Fig.~\ref{tilted-IOs} the vignetting
    functions corresponding to different IO sizes are given.

\begin{figure}
      \resizebox{\hsize}{!}{\includegraphics{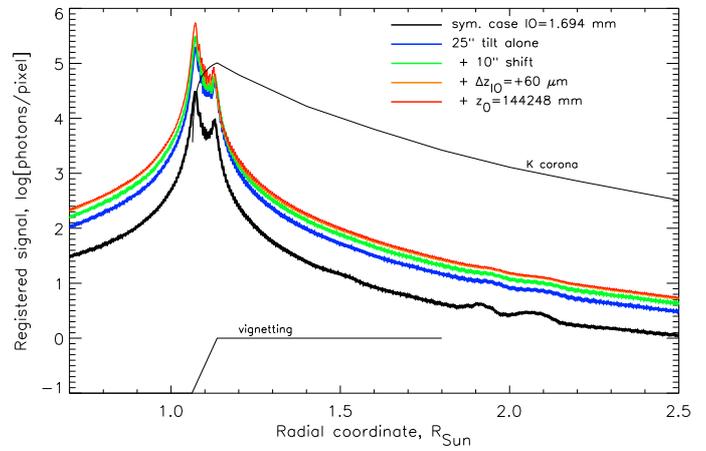}}
      \caption{Simultaneous effect of various misalignments for the IO with $R_{IO}=1.694$~mm. The black curve corresponds to the
	symmetrical case, the colored curves correspond to various combinations of misalignments: 25~arcsec tilt alone (blue); 25~arcsec
	tilt and 10~arcsec shift of the Sun (green); 25~arcsec tilt, 10~arcsec shift of the Sun and 60~$\mu$m displacement of the internal
	occulter (orange);  25~arcsec tilt, 10~arcsec shift of the Sun, 60~$\mu$m displacement of the internal occulter and $\Delta
	z_0=-100$~mm change of the inter-satellite distance (red). The thin curve corresponds to the intensity of the K-corona taking
	into account the vignetting function shown in the bottom part of the plot.}
      \label{simultaneous-misalign}
    \end{figure}

    In order to choose the proper size of the IO, one has to take into account several misalignments simultaneously. An example of the
    combined effect of various misalignments is presented in Fig.~\ref{simultaneous-misalign}. The internal occulter with  
    $R_{IO}=1.694$~mm leads to the diffracted light below the coronal intensity in the unvignetted zone, even with all the misalignments
    combined. It is clear that the major impact is due to the tilt of the telescope. 

  \subsection{IO with apodized profiles}
    \label{apodized}
    \cite{2013A&A...558A.138A} demonstrated that the use of apodized external occulters (i.e. those with variable transmission close to the edge) significantly
    decreases the level of diffraction light at the entrance aperture. Unfortunately, making an apodized external occulter for ASPIICS is a technical challenge.
    However, we demonstrate that apodizing the \emph{internal} occulter may also lead to a significant (although quantitatively different)
    decrease of the diffracted light. 
    
\begin{figure}
      \resizebox{\hsize}{!}{\includegraphics{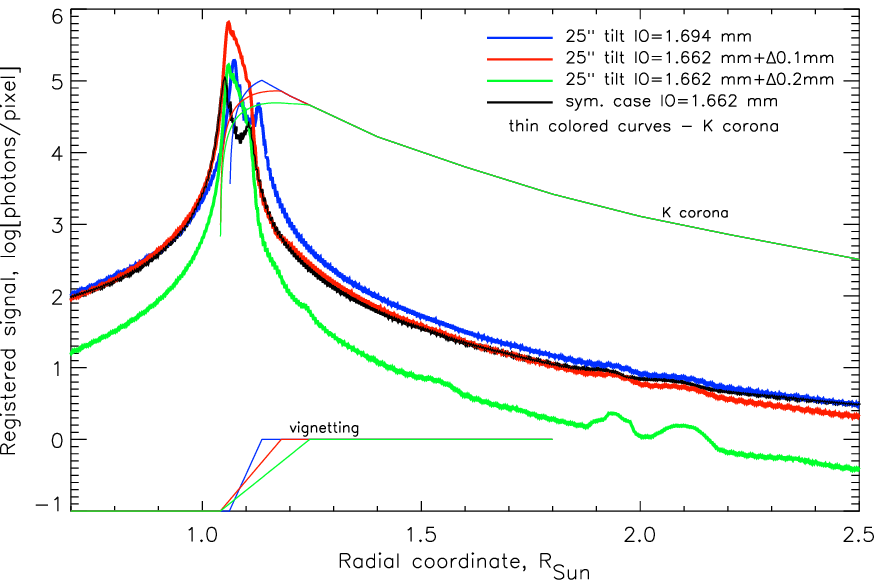}}
      \caption{Comparison of IOs with various apodization for the same tilt of 25~arcsec. The blue curve corresponds to sharp-edge IO with
	$R_{IO}=1.694$~mm, the red curve is for the IO with $R_{IO}=1.662$~mm opaque part and $\Delta=0.1$~mm gradient, the green curve is
	for the with $\Delta=0.2$~mm gradient. Thin colored lines denote coronal signal with corresponding vignetting taken into account and
	vignetting functions (in the bottom part of the plot).}
      \label{radial-step}
    \end{figure}

    In Fig.~\ref{radial-step} we compare the diffracted light for IOs with various apodized profiles -- with the linear gradient of width 0.1 and
    0.2~mm, for the same tilt of 25~arcsec. 
    The thick blue curve corresponds to the sharp-edge IO with $R_{IO}=1.694$~mm. The thick red curve denotes the IO with $R \le 1.662$~mm opaque center and linear gradient
    $\Delta=0.1$~mm apodization, the thick green curve denotes a similar IO but with $\Delta=0.2$~mm apodization. Thin colored lines
    represent coronal signal with additional vignetting by IOs taken into account (colors correspond to the colors of diffraction curves), or the vignetting
    functions (given in the bottom part). 

    IOs with apodization have superior performance with respect to larger sharp-edge IOs. Reduction of the diffracted light is significant: the
    profile of the IO with $\Delta=0.1$~mm apodization and 25~arcsec tilt corresponds to the symmetrical case with $R_{IO}=1.662$~mm
    everywhere except for the innermost region of the corona. The IO with $\Delta=0.2$~mm apodization has the intensity of the diffracted
    light in the inner region almost equal to the symmetrical case, and significantly lower level (factor of 50) of diffracted light at
    heights $>1.1R_\sun$. 
    
    Reduction of the coronal signal occurs in a small region within $1.25R_\sun$, and the number of photons reaching the detector remains
    rather high (see thin curves in Fig.~\ref{radial-step}). For example, for the case of $\Delta=0.2$~mm apodization and during
    $t_{exp}=0.1$~s exposure time, the coronal signal remains at the average level $\sim 3\cdot 10^4$ photons pixel$^{-1}$ for the range of
    coronal heights $1.1-1.25R_\sun$. Such a photon flux allows reliable registration of the signal. 

\section{Discussion}
  \label{discussion-sec}
  \subsection{Comparison of various misalignments}
    The performed computations show that the impact of various misalignments is considerably different for each type. For example, shifts
    of the EO and the coronagraph may have both severe and negligible impacts on the diffracted light. Non-symmetrical umbra/penumbra pattern on
    the entrance aperture not necessarily produces a severe impact (e.g. solar shift). Alternatively, symmetrical umbra/penumbra pattern on
    the entrance aperture does not guarantee good diffracted light performance (e.g. tilt of the coronagraph). Placing the EO either closer
    to the coronagraph or further away does not result in a significant increase of the diffracted light for the expected displacement.
    However, the displacement of the IO from its nominal position along the $z$-axis (which may resemble the displacement of the EO) either worsens or
    improves the  performance depending on the position in the field of view.

  \subsection{Using apodized IOs}
    As we show in Sect.~\ref{apodized}, the IO with 0.2~mm apodization significantly reduces the level of diffracted light in tilted cases.
    Obviously, the performance of the apodized IO will be even better at smaller tilts or in symmetrical cases. 
    
    The reduction of the coronal signal remains reasonable, as the inner corona is bright enough. Furthermore, the reduction of dynamic range
    produced by apodized IOs may further improve the performance of the coronagraph by diminishing the level of ghosts and scattered light in the
    instrument.

    Using a smaller opaque region in apodized IOs potentially provides additional advantages, as it reduces $v_{min}$ -- the height at which
    vignetting starts to disappear. In our case apodized IOs have $v_{min}=1.078R_\sun$, whereas a larger IO with $R=1.694$~mm has
    $v_{min}=1.099R_\sun$. From this point view, even the IO with apodization $\Delta=0.1$~mm may have an advantage over larger
    sharp-edge IOs with the same level of diffraction.

    From a technological point of view, manufacturing of apodized IOs may be easier than that of sharp-edge IOs: producing a very sharp edge
    may not be possible, and the remaining edge irregularities may additionally increase diffraction. In any case, accurate measurements of
    the IO geometry and transmission are of a vital importance for processing of the registered coronal images. 

    Potential phase shifts of the wavefront due to propagation through the apodized IOs still need to be investigated.

  \subsection{Dependence on the wavelength}
    \label{wavelength}
    ASPIICS is equipped with a filter wheel with 6 positions, which can be switched between individual exposures \citep{Renotte2016}. There are 3 passbands:
    ``wide-band'' 535--565 nm, narrow-band 530.4 nm for the \ion{Fe}{XIV} line, and narrow-band 587.7~nm for the \ion{He}{I} line. Three
    additional positions are equipped with polarizers (rotated by $60^{\circ}$ with respect to each other) together with the wide-band filter.
    There are various effects that influence the performance of ASPIICS in various passbands. First, chromatic aberrations of the primary
    objective and the rest of the optics very slightly modify the point-spread-function of the telescope. Second, it is essential that in
    the narrow-band filters the observed coronal structures will be bright in both spectral lines \emph{and} continuum, thus the relative effect of the diffraction
    will be less important. 
    
    For the diffraction-related issues, we use the following reasoning: due to the property of diffraction discussed in
    Sect.~\ref{properties}, the diffraction image in the $O'$ plane looks like a bright ring with the angular size of the EO. Obviously, this
    property will be valid for \emph{any} wavelength, despite the fact that the diffraction pattern from an individual plane-parallel wave
    $\Psi_{A\vec k}$ depends on $\lambda$ and the diffraction at the entrance aperture also depends on $\lambda$. As a consequence,
    occulting the beam by the IO will occur with almost the same efficiency, regarding of the particular wavelength. Additional effects, like
    different radial decrease of the intensity of the bright ring (the radial size of the Airy spot scales as $1.22 \lambda D/f$), may still be
    present, but the major effect of blocking the diffracted light by the IO will be the same. 

    Thus we conclude that for the multi-wavelength observations diffraction effects do not change. A more detailed analysis of the spectral
    dependence of the diffraction will be carried out in the future. 

\section{Conclusions}
  \label{conclusion-sec}
  We analyzed different types of misalignments in the externally occulted ASPIICS coronagraph onboard the PROBA-3 mission, which has a very
  small overoccultion of the solar disk. We considered impacts of misalignments from the physical point of view and their computational
  realizations. We computed the resulting pattern of the diffracted light in the detector plane and compared it with the coronal
  intensity. 
  
  Our computations of the diffracted light can be applied to any coronagraph with arbitrary geometry. However, for the case of ASPIICS we show
  an exceptional importance of precise co-alignment. The most important misalignment is the tilt of the coronagraph. Special care should be taken to co-align
  the external and internal occulters. The true optical axis of the instrument (through the center of the IO and the entrance aperture) should
  be pointed to the center of EO as close as possible. The margin for the acceptable tilts $\sim 25$~arcsec is determined by oversizing the IO over
  the EO; the margin can be increased in case of an apodized IO. Impacts of other misalignments are significantly smaller. Since the tilt
  of the coronagraph can be potentially corrected in flight by correcting the attitude of the coronagraph satellite, there is a possibility to
  reduce the size of the IO and, as a consequence, to reduce the minimal observed height of the corona. 

  We found that apodized IOs have very good diffracted light rejection performance, and especially are resistant to tilts.

\begin{acknowledgements}
  We acknowledge support from the Belgian Federal Science Policy Office through the ESA -- PRODEX programme (grant No.~4000117262). Authors
  are grateful to Dr. Anton Reva for his help during the preparation of the manuscript. We are grateful to anonymous referee for numerous
  suggestions, which allowed to improve the paper.  
\end{acknowledgements}

\bibliographystyle{aa}
\bibliography{shestov_zhukov}

\appendix
\section{Details of the mathematical approach}
    \label{method}
    In our analysis we follow the method of \citet{2013A&A...558A.138A} and RR17. First we consider the propagation of a wave
    $\Psi$ through an optical system \citep{goodman2005introduction}. After propagation through an aperture $A$, at the distance $z$ the wave
    amplitude $\Psi_z$ is calculated as:
    \begin{equation}
	    \Psi_z (x,y) = \frac{\exp \left( \frac{i\pi (x^2+y^2)}{\lambda z} \right)}{i\lambda z} \cdot \mathcal F  
	    \left\{ A \cdot \Psi(\xi,\eta) \cdot \exp \left(\frac{i \pi (\xi^2+\eta^2)}{\lambda z} \right) \right\},
    \end{equation}
    where $(x,y)$ -- coordinates in the observer plane, $\mathcal F$ -- Fourier transform, $(\xi,\eta)$ -- coordinates in the aperture
    plane, $\Psi(\xi,\eta)$ -- amplitude of the initial wave, $\lambda$ -- wavelength, $A = A(\xi,\eta)$ -- aperture function that denotes transparency at point $(\xi, \eta)$. 

    If one places a convergent lens with the focal distance $f$ immediately after the aperture, one must add an additional term $\exp \left( -
    \frac{i \pi (\xi^2+\eta^2)}{\lambda f} \right)$ inside the $\mathcal F$ argument. In the case $z$ and $f$ coincide, one can obtain a
    well-known expression for the focal plane of the lens:
    \begin{equation}
      \Psi_z (x,y) = \frac{\exp \left( \frac{ i \pi (x^2+y^2)}{\lambda z} \right)}{i\lambda z} \cdot \mathcal F \left\{ A \cdot \Psi(\xi,\eta) \right\}.
    \end{equation}

    Keeping this in mind, we write the expression for the field amplitude in $O'$:
    \begin{multline}
      \Psi_{O'}= \frac{ \exp \left( \frac{ i \pi r^2}{\lambda z_1} \right) }{i \lambda z_1} \cdot \mathcal F \left\{ A_A \cdot \Psi_A \cdot \exp \left( \frac{
	i \pi r^2}{\lambda z_1} \right) \cdot \exp \left( - \frac{i \pi r^2}{\lambda f} \right) \right\} = \\
	= \frac{ \exp \left( \frac{i \pi r^2}{\lambda z_1} \right) }{i \lambda z_1} \cdot \mathcal F_{O'},
    \end{multline}
    for the field amplitude in $C$:
    \begin{multline}
      \Psi_C = \frac{ \exp \left( \frac{i \pi r^2}{\lambda d} \right) }{i \lambda d} \cdot \mathcal F \left\{ A_{O'} \cdot \Psi_{O'} \cdot \exp
      \left( \frac{i \pi r^2}{\lambda d} \right) \cdot \exp \left( - \frac{i \pi r^2}{\lambda f_{L2}} \right) \right\} = \\
      = \frac{ \exp \left( \frac{i \pi r^2}{\lambda d} \right) }{(i \lambda)^2 z_1 d} \cdot \mathcal F_C,
    \end{multline}
    and finally for the field amplitude in $D$:
    \begin{multline}
      \Psi_D = \frac{ \exp \left( \frac{i \pi r^2}{\lambda l} \right) }{i \lambda l} \cdot \mathcal F \left\{ A_C \cdot \Psi_C \cdot \exp
	\left( \frac{i \pi r^2}{\lambda l} \right) \cdot \exp \left( - \frac{ i \pi r^2}{\lambda f_{L3}} \right) \right\} = \\
	= \frac{ \exp \left( \frac{i \pi r^2}{\lambda l} \right) }{ (i \lambda)^3 z_1 d l} \cdot \mathcal F \left\{ A_C \cdot \mathcal F_C
	  \cdot \exp \left[ \frac{ i \pi r^2}{\lambda} \left( \frac1d + \frac1l - \frac1{f_{L3}} \right) \right] \right\} 
    \end{multline}

    One can further simplify the expressions for the case of the optical layout corresponding to the initial design (i.e. taking into
    account $1/z_0 + 1/z_1 = 1/f$, $d=z_1$, $f_{L2}=z_1/2$, $l=f$, and $f_{L3}=z_1/2$), but for consideration of displaced optical
    components one should take specific values. 
   
    We note that arguments in each Fourier transform are taken at a corresponding plane and the
    coordinates and linear scales are different in each case. To perform numerical modeling of the Fourier propagation,
    we substitute 2D functions $\Psi_A$, $A_A$, $\Psi_{O'}$, $A_{O'}$, $\exp \left( \frac{i \pi r^2}{\lambda z_1}
    \right)$, etc. by corresponding 2D arrays.

    Now we consider the diffraction of a plane-parallel wave on an infinitely thin circular occulter. The problem was considered by
    \citet{2013A&A...558A.138A}, with the result that the amplitude can be expressed via the amplitude of the initially co-axial wave $\Psi_{A00}$ (see his Eq.~(5)):
    \begin{equation}
	\Psi_{A \alpha \beta} (x,y) = 
	  T_{\alpha \beta}(x,y) \cdot \Gamma_{\alpha \beta}(x,y) \cdot \Psi_{A00} (x+\alpha z_0, y+\beta z_0), 
	\label{PsiA}
    \end{equation}
    where $\Psi_{A00} (\xi,\eta)$ is the amplitude of the initially co-axial wave, and 
    \begin{gather}
      T_{\alpha \beta} (x,y) = \exp \left( -2 \pi i \frac{\alpha x + \beta y}{\lambda} \right) \qquad \mathrm{(Tilt)} \\
      \Gamma_{\alpha \beta} (x,y) = \exp \left( - \pi i \frac{ (\alpha^2 + \beta^2) z_0}{\lambda} \right) \qquad \mathrm{(Offset)}.
    \end{gather}
    $\Psi_{A00} (\xi,\eta)$ is calculated using the Fourier-Hankel transform:
    \begin{equation}
      \Psi_{A00} (\xi,\eta) = 1- \frac{\varphi_{z0}(r)}{i \lambda z_0} \int_0^R 2 \pi \rho \exp
      \left( i \pi \frac{\rho^2}{\lambda z_0} \right) J_0 \left( 2\pi \frac{r \rho}{\lambda z_0} \right) \, d\rho, 
    \end{equation}
    where $r=\sqrt{\xi^2 + \eta^2}$ is the radial coordinate in the $A$ plane, $\varphi_{z0} = \exp (i\pi r^2/ \lambda
    z_0)$, and $J_0(r)$ is the Bessel function of the first kind. The function $\Psi_{A00}$ is circularly symmetrical
    and is calculated initially as a 1D-array.

\section{Numerical issues}
  \label{numerical}
  Calculation of $\Psi_{A00}$ is computationally very expensive: computation of $\Psi_{A00}$ with the spatial sampling of 3 $\mu$m takes almost 10
  hours using IDL language on a PC with Intel Core i5 6200 processor. So we computed the function once for the nominal $z_0=144 \, 348$~mm,
  and displaced $\Delta z_0 = +15$, $-15$~mm, and saved them into separate files ($\lambda=550$~nm and $R_{EO}=710$~mm were used). Computation
  of $\Psi_{A\vec k}$ from $\Psi_{A00}$ takes 2--3 seconds, and computation of $\Psi_{O'}$, $\Psi_C$, $\Psi_D$ takes $\sim 15$~seconds for
  $4k \times 4k$ arrays. The difficulty is due to the necessity of high spatial sampling of the Sun: if we want to sample e.g. by
  $N=50$~points in $\rho$ and $M=400$~points in $\varphi$, then the total integration time will be $\tau=20NM$~seconds, or $\sim 110$~hours.
  Such a long time is not acceptable for performing a parametric study. 

  For the case of the perfect symmetry, RR17 used a method that resembles our approach described in Appendix~\ref{sampling-sec}: $|\Psi_{D
  \vec k}|^2$ is calculated for $\varphi=0$, and after that the obtained image is ``blurred'' over $2\pi$ in the polar
  coordinate (due to the circular symmetry of the source -- the Sun). Since in the present analysis we are interested in non-symmetrical
  configurations, a similar approach can not be used for the full Sun.  
  

  In order to speed-up the computations we developed an MPI-Fortran version of the full sampling model. Using of Fortran with \verb{fftw{
  \citep{Frigo2005} library significantly increases the computation speed: 3 sequential FFT transforms now take only $\sim 5$~seconds.
  Furthermore, the algorithm can be easily parallelized: an individual process computes a given subset of incident waves, and after that all
  the $I_D$ are summed. As a result, most of the computations reported here were carried out using full-sampling approach with $N=50$, and
  $M=400$~points. A single computation with 128 MPI processes on a cluster with Intel Xeon E5-2680 processors took on average around
  $~20$~minutes. Necessity for parallelization on our own cluster did not allow us using general purpose optical
  libraries like PROPER~\citep{Krist2007}\footnote{Available freely from SourceForge: http://proper-library.sourceforge.net}.

\section{Influence of spatial sampling}
  \label{sampling-sec}
  We analyzed the influence of spatial sampling of the Sun and sizes of the arrays representing $\Psi_A$, $\Psi_{O'}$, and $\Psi_D$ on our
  results. 

  Most of the images presented in this work were computed with sampling $M=400$ points in $\varphi$. Such a sampling reveals
  modulation in the final images along the perimeter of the bright annulus (imagine 400 images from Fig.~\ref{property}d rotated
  and stacked together). In order to simulate a higher sampling in $M$, we used the following method: we substitute each image $I_D$
  with an image ``blurred'' in polar coordinate $I_D^\prime=1/W \sum_0^{W-1} rot(I_D,\alpha)$, where $I_D$ is a low-sampled image,
  $\alpha \in [0 ..  2\pi/400)$ are equally spaced $W$ points, $rot$ is a procedure for rotation of an image $I_D$ by an angle $\alpha$. We
  applied this additional step every time in a separate IDL procedure.  The obtained ``blurred'' images coincided with unrotated images
  $I_D$ computed with $M=1600$ sampling, which have no modulation in $\varphi$. The small step size of the initial sampling $2\pi/400$
  results in obtained $I_D^\prime$ being smooth in the polar coordinate. Our method resembles the method used by RR17 for simplified
  computations of symmetrical configurations (see their Eq.~12 and corresponding description).
  
\begin{figure}
    \resizebox{\hsize}{!}{\includegraphics{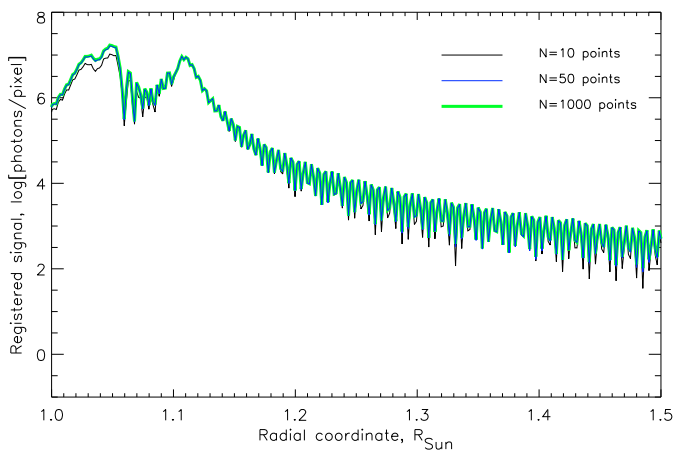}}
    \caption{Comparison of various sampling of the Sun over the radius. The black curve corresponds to $N=10$ point, the thin blue curve is to 
      $N=50$ points, and thick green curve is to $N=1000$ points.}
    \label{compare-sampling}
  \end{figure}

  The difference between images with different sampling in $\rho$ (with $N=10$, $N=50$, $N=200$, and $N=1000$ points) could be barely
  noticed (Fig.~\ref{compare-sampling}) so we used $N=50$ sampling for the analysis.
 
  We also tried to use ``Cartesian'' strategy to sample the solar disk, sampling points uniformly along $x$- and $y$-axes. We expected that
  the ``polar'' strategy (points are sampled uniformly along $\rho$ and $\varphi$) tends to over-concentrate points at low $\rho$ values
  and assigns them low importance due to small $\mbox{d}S$. The ``Cartesian'' strategy with $\approx 20\,000$ points (the same number as in
  the ``polar'' case) provided smoother images in $D$, i.e. with a smaller but still noticeable modulation in $\varphi$.  However, such a
  strategy does not allow to apply additional image ``blurring'', thus the advantage of the ``Cartesian'' strategy over the ``polar'' one
  disappears. 
  
  Change of the size of arrays from $4k \times 4k$ to $8k \times 8k$ also did not produce an significant change. However, increases in
  computation time and volume of the data were significant.

  The most significant influence on diffraction is the spatial sampling of the $\Psi_{A00}$ function. Here we used 3 $\mu$m spatial sampling,
  which is $\sim 6$ times smaller than 17~$\mu$m the spatial scale in the $A$ plane (i.e. scale size of the $\Psi_A$ array).  Reduction of
  $\Psi_{A00}$ sampling to 7 $\mu$m results in the development of ``ghost'' arches in the $D$ plane, whereas at 3 $\mu$m these arches are almost
  invisible. The arches are weak even in case of the 7 $\mu$m sampling, and are most pronounced in the outer regions of the field of view, where the
  diffracted signal is low. An additional improvement of $\Psi_{A00}$ sampling may further modify the level of the diffracted light in outer
  regions. However, we limit our analysis to 3 $\mu$m sampling due to already a low level of diffraction at high altitudes and
  presence of additional factors that may degrade the images (primarily ghost images produced by unwanted reflections from lens surfaces).

\end{document}